# G1-Renewal Process as Repairable System Model

M.P. Kaminskiy[‡] and V.V. Krivtsov[†]

[‡]University of Maryland, College Park, USA, [†]Ford Motor Company, Dearborn, USA

**Abstract**

This paper considers a point process model with a monotonically decreasing or increasing ROCOF and the underlying distributions from the location-scale family, known as the *geometric process* (Lam, 1988). In terms of repairable system reliability analysis, the process is capable of modeling various restoration types including "better–than–new", i.e., the one *not* covered by the popular G-Renewal model (Kijima & Sumita, 1986). The distinctive property of the process is that the times between successive events are obtained from the underlying distributions as the scale parameter of each is monotonically decreasing or increasing. The paper discusses properties and maximum likelihood estimation of the model for the case of the Exponential and Weibull underlying distributions.

*Acronyms*:

| | |
|---|---|
| CDF | cumulative distribution function |
| CIF | cumulative intensity function |
| GPR | generalized renewal process |
| HPP | homogeneous Poison process |
| IID | independent and identically distributed |
| MLE | maximum likelihood estimation |
| NHPP | non-homogeneous Poison process |
| ORP | ordinary renewal process |
| PDF | probability density function |
| ROCOF | rate of occurrence of failures |

## 1. Introduction

In repairable system reliability analysis, if upon a failure, a system is restored to as "good-as-new" condition and the time between failures can be treated as an independent and identically distributed (IID) random variable, then the failure occurrence can be modeled by the *Ordinary Renewal Process* (ORP).

If upon a failure the system is restored to the "same-as-old" condition, then the appropriate model to describe the failure occurrence is the *Non-Homogeneous Poisson Process* (NHPP). The time between consecutive failures, in this case, is not an IID random variable. In a sense, the NHPP can be viewed as a renewal process with the "same-as-old" repair assumption (Krivtsov, 2007). An important particular case of both ORP and NHPP is the *Homogeneous Poisson Process* (HPP), whose underlying failure times are distributed exponentially.

It is clear that even though attractive mathematically, the "good-as-new" and "same-as-old" repair assumptions are often exceptions rather than the rule, from the standpoint of practical reliability engineering. Generally, they could be treated as the "limiting" conditions to which a system could be restored. In reality, after the repair, the system is likely to find itself *between* the two conditions. Of great interest, therefore, is modeling other repair assumptions such as the intermediate "better-than-old-but-worse-than-new".

An early approach to cover more than one repair assumption within the same probabilistic model is discussed in (Brown and Proschan, 1982). This method assumes that upon a failure, a repair action restores the system to the "good-as-new" condition with probability of *p(t)*, or the "same-as-old" condition with probability of 1-*p(t)*, where *t* is the age of the system at failure.

A more general model is the so-called *G-Renewal* process (Kijima M and Sumita, 1986), which treats ORP and NHPP as special cases. The GRP is introduced using the notion of *virtual age*:

$$A_n = qS_n,$$

where $A_n$ and $S_n$ is the system's *virtual age* before and after the *n*-th repair, respectively, and *q* is the *restoration* (or *repair effectiveness*) *factor*.

It is clear that for *q* = 0, the age of the system after the repair is "re-set" to zero, which corresponds to the "good-as new" repair assumption and represents the ORP. With *q* = 1, the system is restored to the "same-as-old" condition, which is the case of the NHPP. The case of 0 < *q* < 1 corresponds to the intermediate "better-than-old-but-worse-than-new" repair assumption. Finally, with *q* > 1, the virtual age is $A_n > S_n$, so that the repair damages the system to a higher degree than it was just before the respective failure, which corresponds to the "worse-than-old" repair assumption.

One limitation of the GRP model is its inability to model a "better than new" restoration, for which the need arises in some practical applications, e.g. *reliability growth modeling* (Crow, 1982). The considered below geometric process (Lam, 1988, 2009) overcomes this particular drawback.

## 2. Geometric process: probabilistic model

The location-scale family of underlying distributions is considered. After each *i*-th failure (*i* = 1, 2,...), the system is restored (damaged) in such a way that its scale parameter *α* is changed to $\alpha(1+q)^{i-1}$, where $q$ is the restoration (damage) parameter, $-1 < q < \infty$, so that for the time to the first failure $i = 1$, for the time between first and second failure $i = 2$, and so on. This transformation of the scale parameter is similar to the one used in the well-known accelerated life time model (Cox & Oaks, 1984; Nelson, 1990). To an extent, the suggested model makes more physical (reliability) sense than the respective NHPP model in terms of restoration assumption (i.e., "same-as-old" assumption). If $q = 0$, the process coincides with the ordinary Renewal process. If $q > 0$, the introduced process is obviously an improving one, and if $q < 0$, the process is aging (deteriorating). Table 1 shows multiplier $(1+q)^{i-1}$ to the scale parameter of the underlying distributions of the times between consecutive events for some values of $q$.

*Table 1. Multiplier $(1+q)^{i-1}$ to the scale parameter of the underlying Distributions of the times between successive events for some values of q.*

| Event, $i$ | $q = 0.1$ | $q = -0.1$ | $q = 0.2$ | $q = -0.2$ |
|---|---|---|---|---|
| 1 | 1.000 | 1.000 | 1.000 | 1.000 |
| 2 | 1.100 | 0.900 | 1.200 | 0.800 |
| 3 | 1.210 | 0.810 | 1.440 | 0.640 |
| 4 | 1.331 | 0.729 | 1.728 | 0.512 |
| 5 | 1.464 | 0.656 | 2.074 | 0.410 |
| 6 | 1.611 | 0.590 | 2.488 | 0.328 |
| 7 | 1.772 | 0.531 | 2.986 | 0.262 |
| 8 | 1.949 | 0.478 | 3.583 | 0.210 |
| 9 | 2.144 | 0.430 | 4.300 | 0.168 |
| 10 | 2.358 | 0.387 | 5.160 | 0.134 |

In the given context, we suggest calling the considered geometric point process as the *G1-Renewal Process* due to a certain similarity to the G-Renewal Process introduced earlier by Kijima and Sumita (1986). Again, by analogy with G-Renewal Equation, the equation for the cumulative intensity function (CIF) of the G1-Renewal Process will be correspondingly called the *G1-Renewal equation*. It should be noted that the process does not have an established name, e.g., Wang and Pham (2006) call this process a *quasi-renewal process*.

*2.1. G1-Renewal Equation*

The location-scale distribution for a continuous random variable (r. v.) $t$ is defined as having the cumulative distribution function (CDF) in the following form:

$$F(t) = F\left(\frac{t-u}{\alpha}\right) \tag{1}$$

The respective probability density function is

$$f(t) = \frac{1}{\alpha}\left(\frac{t-u}{\alpha}\right) \tag{2}$$

The time to the $n$th failure $T_n$ is given by

$$T_n = X_1 + X_2 + \ldots + \ldots X_n = \sum_{i=1}^{n} X_i \tag{3}$$

where $X_i$ ($i = 1, 2, \ldots, n$) are independent r.v., which, in the framework of the G1-Renewal Process, are distributed according to the following cumulative distribution function (CDF)

$$F_i(X_i) = F\left(\frac{X_i - u}{\alpha(1+q)^{i-1}}\right), i = 1, 2, \ldots, n \tag{4}$$

The distribution of the time to the $n$-th failure $T_n$ is difficult to find as a closed-form expression, even in the case of the ordinary renewal process, i.e. when $q = 0$ (except for the exponential and Gamma distribution among the popular lifetime distributions). Note that in the process considered, contrary to the ordinary renewal process, the $X_i$ 's are not identically distributed.

The equation for the *cumulative intensity function* (CIF), also known as the *g-renewal equation* of the process can be found as

$$W(t) = \sum_{k=1}^{\infty} F^{(k)}(t) \tag{5}$$

where $F^{(k)}(t)$ is $k$-fold convolution of the cumulative distribution functions (4). Note that $F^{(k)}(t) = Pr(T_k < t)$. The respective *rate of occurrence of failures* (ROCOF) can be found using its definition as

$$w(t) = \frac{dW(t)}{dt} = \sum_{k=1}^{\infty} f^{(k)}(t) \tag{6}$$

*2.2. G1-Renewal Process with Exponential Underlying Distribution*

The process with exponential underlying distribution is considered. The time to the first failure has the exponential distribution with PDF

$$f_1(t) = \frac{1}{\alpha}\exp\left(-\frac{t}{\alpha}\right) \tag{7}$$

According to (4), the time between the first and the second failures has the following PDF

$$f_2(t) = \frac{1}{\alpha(1+q)}\exp\left(-\frac{t}{\alpha(1+q)}\right), \tag{8}$$

Correspondingly, the time between the (i-1)th and the i-th failures has the following PDF

$$f_i(t) = \frac{1}{\alpha(1+q)^{i-1}}\exp\left(-\frac{t}{\alpha(1+q)^{i-1}}\right) \tag{9}$$

The convolution of $f_1(t)$ and $f_2(t)$, i.e., $f_1(t)*f_2(t)$ can be found as

$$\begin{aligned}
f_1(t)*f_2(t) &\equiv f^{(2)}(t) = \int_0^t f_1(t-x)f_2(x)dx = \int_0^t f_2(t-x)f_1(x)dx \\
&= \frac{1}{\alpha^2(1+q)}\int_0^t \exp\left(-\frac{t-x}{\alpha}\right)\exp\left(-\frac{x}{\alpha(1+q)}\right)dx \\
&= \frac{1}{\alpha^2(1+q)}\exp\left(-\frac{t}{a}\right)\int_0^t \exp\left(\frac{xq}{\alpha(1+q)}\right)dx \\
&= \frac{1}{aq}\exp\left(-\frac{t}{a}\right)\left[\exp\left(\frac{tq}{a(1+q)}\right)-1\right]
\end{aligned} \tag{10}$$

It can be shown that the Laplace transform of the PDF of time to the $i^{th}$ failure $f_i(t)$ is given by

$$f_i^*(s) = \frac{1}{sa(1+q)^{i-1}+1} \tag{11}$$

Based on (11), the Laplace transform of the convolution $f^{(k)}(t)$ can be found as

$$f^{*(k)}(s) = \prod_{i=1}^k \frac{1}{sa(1+q)^{i-1}+1} \tag{12}$$

The inverse of (12) is not available in a closed form, which is why we default to obtaining the CIF for G1-Renewal Process via Monte Carlo simulation – similar to the solution of the G-Renewal equation we suggested in (Kaminskiy & Krivtsov, 1998).

Figures 1 and 2 show the CIF's of the G1-Renewal Process with underlying exponential distribution. It is interesting to note that in the context of the G1-Renewal, the underlying exponential distribution provides a high flexibility in modeling both *improving* and *deteriorating* processes – contrary to the HPP.

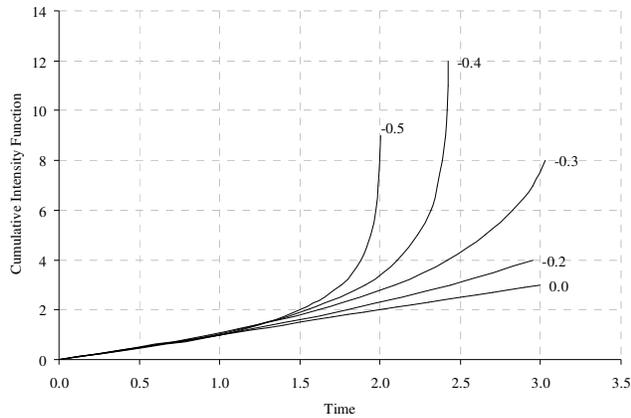 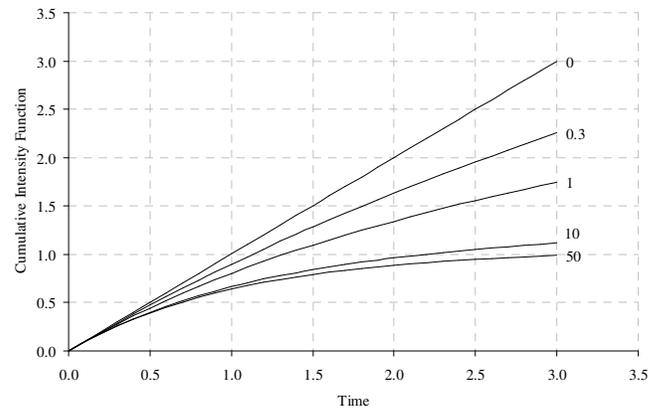

*Fig. 1. CIF of the G1-Renewal Process with underlying exponential distribution, scale parameter of 1 and various negative values of q.*

*Fig. 2. CIF of the G1-Renewal Process with underlying exponential distribution, scale parameter of 1 and various positive values of q.*

*2.3. G1-Renewal Process with Weibull Underlying Distribution*

Figures 3 and 4 show the CIF's of the G1-Renewal Process with the positive restoration parameter and the underlying Weibull distribution with the scale parameter of 1 and the increasing and decreasing hazard functions, respectively.

The concavity of the CIF for t < ~0.7 in Figure 3 might be related to the increasing hazard function of the underlying distribution. The subsequent convexity of the CIF for t > 0.7 might be explained by the positive restoration parameter, which corresponds to the improving G1R process. The overall convexity of the CIF in Figure 4 might be explained by the decreasing hazard function of the underlying distribution and the positive restoration parameter, which corresponds to the improving G1R process.

The concavity of the CIF in Figure 5 might be explained by the increasing hazard function of the time-to-first-failure distribution and a negative restoration parameter, which corresponds to the deteriorating G1R process. The relative "linearity" of the CIF in Figure 6 might be explained by the decreasing hazard function of the underlying distribution, which is partially "compensated" by the negative restoration parameter of the G1R process.

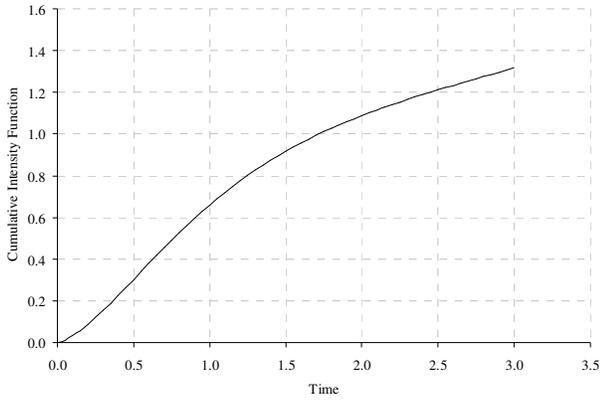

*Fig. 3. CIF of the G1-Renewal Process with underlying Weibull distribution, scale parameter of 1, shape parameter of 1.5 and restoration parameter of 3.*

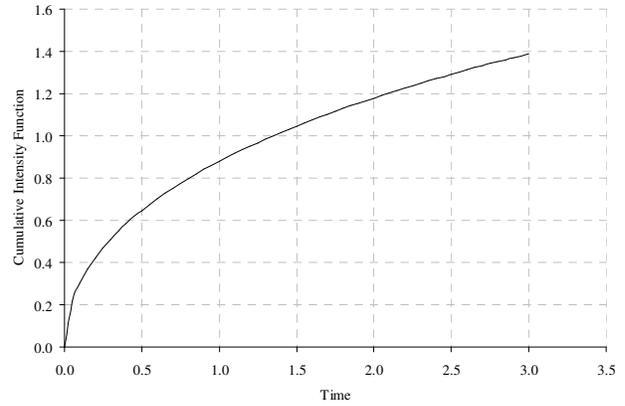

*Fig. 4. CIF of the G1-Renewal Process with underlying Weibull distribution, scale parameter of 1, shape parameter of 0.5 and restoration parameter of 3.*

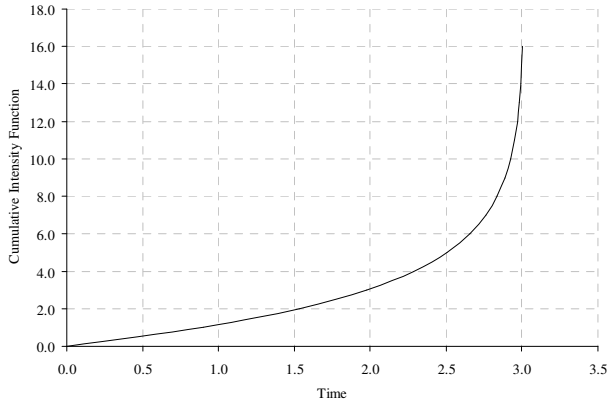

*Fig. 5. CIF of the G1-Renewal Process with underlying Weibull distribution, scale parameter of 1, shape parameter of 1.5 and restoration parameter of -0.3.*

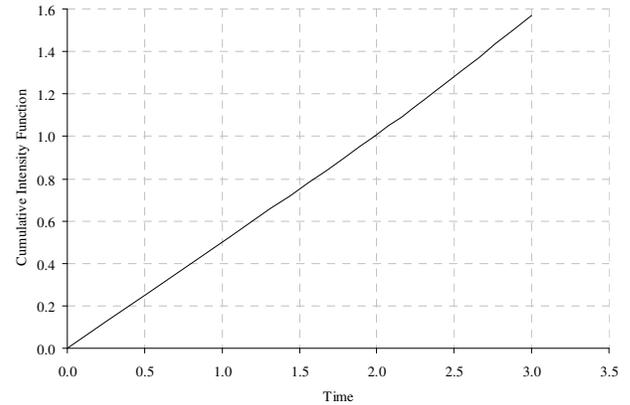

*Fig. 6. CIF of the G1-Renewal Process with underlying Weibull distribution, scale parameter of 1, shape parameter of 0.5 and restoration parameter of -0.3.*

### 3. G1-renewal process: maximum likelihood estimation

*3.1. Data*

Let $t_1$ be time to the first failure, $t_2$ be the time between the first failure and the second failure, so that $t_n$ is the time between the $(n-1)$th failure and the last $n$th failure. The test (observation) is terminated at the time $t = t_n$.

*3.2. G1-Renewal Equation with Exponential Underlying Distribution*

For the underlying distribution (7) the likelihood function can be written as follows:

$$L(a,q) = \left[\frac{1}{a}\exp\left(\frac{-t_1}{a}\right)\right]\left[\frac{1}{a(1+q)}\exp\left(\frac{-t_2}{a(1+q)}\right)\right]\left[\frac{1}{a(1+q)^2}\exp\left(\frac{-t_2}{a(1+q)^2}\right)\right]\cdots$$
$$\left[\frac{1}{a(1+q)^{n-1}}\exp\left(\frac{-t_2}{a(1+q)^{n-1}}\right)\right]$$

Taking logarithms of the function and differentiating with respect to $a$ and $q$ one gets

$$\begin{cases} \dfrac{d(\ln(L(a,q)))}{da} = \dfrac{1}{a}\sum_{i=1}^{n}\dfrac{t_i}{(1+q)^{i-1}} - n = 0 \\ \dfrac{d(\ln(L(a,q)))}{dq} = \dfrac{1}{1+q} + \dfrac{n(1-n)}{2} - \dfrac{1}{a}\sum_{i=1}^{n}\dfrac{(1-i)t_i}{(1+q)^i} = 0 \end{cases} \quad (13)$$

System of equations (13) can be solved numerically.

*3.3. G1-Renewal Equation with Weibull Underlying Distribution*

As in the previous case, the same type of failure-terminated data are considered. The PDF of the underlying (time to the first failure) Weibull distribution is

$$f_1(t) = \frac{\beta}{\alpha}\left(\frac{t}{\alpha}\right)^{\beta-1}\exp\left[-\left(\frac{t}{\alpha}\right)^{\beta}\right] \quad \alpha, \beta > 0, t \geq 0 \quad (14)$$

For the above underlying distribution the likelihood function is

$$L(\alpha,\beta,q) = \frac{\beta}{\alpha}\left(\frac{t_1}{\alpha}\right)^{\beta-1}\exp\left[-\left(\frac{t_1}{\alpha}\right)^{\beta}\right] X \frac{\beta}{\alpha(1+q)}\left(\frac{t_2}{\alpha(1+q)}\right)^{\beta-1}\exp\left[-\left(\frac{t_2}{\alpha(1+q)}\right)^{\beta}\right] \bullet$$
$$\frac{\beta}{\alpha(1+q)^2}\left(\frac{t_3}{\alpha(1+q)^2}\right)^{\beta-1}\exp\left[-\left(\frac{t_3}{\alpha(1+q)^2}\right)^{\beta}\right]\cdots \frac{\beta}{\alpha(1+q)^{n-1}}\left(\frac{t_n}{\alpha(1+q)^{n-1}}\right)^{\beta-1}\exp\left[-\left(\frac{t_n}{\alpha(1+q)^{n-1}}\right)^{\beta}\right] \quad (15)$$

Taking the logarithm of this likelihood function one gets:

$$\ln(L(\alpha,\beta,q)) = \ln\left(\frac{\beta}{\alpha}\right) + (\beta-1)\ln\left(\frac{t_1}{\alpha}\right) - \left(\frac{t_1}{\alpha}\right)^{\beta} + \ln\left(\frac{\beta}{\alpha(1+q)}\right) + (\beta-1)\ln\left(\frac{t_2}{\alpha(1+q)}\right) - \left(\frac{t_2}{\alpha(1+q)}\right)^{\beta} +$$
$$+ \ln\left(\frac{\beta}{\alpha(1+q)^2}\right) + (\beta-1)\ln\left(\frac{t_3}{\alpha(1+q)^2}\right) - \left(\frac{t_3}{\alpha(1+q)^2}\right)^{\beta} + \ldots + \ln\left(\frac{\beta}{\alpha(1+q)^{n-1}}\right) + (\beta-1)\ln\left(\frac{t_n}{\alpha(1+q)^{n-1}}\right) - \left(\frac{t_n}{\alpha(1+q)^{n-1}}\right)^{\beta} \quad (16)$$

Differentiating this function with respect to $\alpha$, $\beta$ and $q$, and equating the derivatives to zero one gets:

$$\frac{d\ln(L(\alpha,\beta,q))}{d\alpha} = -\frac{\beta}{\alpha} + \frac{\beta}{\alpha}\left(\frac{t_1}{\alpha}\right)^{\beta} - \frac{\beta}{\alpha} + \frac{\beta}{\alpha}\left(\frac{t_2}{\alpha(1+q)}\right)^{\beta} - \frac{\beta}{\alpha} + \frac{\beta}{\alpha}\left(\frac{t_3}{\alpha(1+q)^2}\right)^{\beta}\ldots - \frac{\beta}{\alpha} + \frac{\beta}{\alpha}\left(\frac{t_n}{\alpha(1+q)^{n-1}}\right)^{\beta}$$
$$= -\frac{n\beta}{\alpha} + \frac{\beta}{\alpha}\sum_{i=1}^{n}\left(\frac{t_i}{\alpha(1+q)^{i-1}}\right)^{\beta} = -n + a^{-\beta}\sum_{i=1}^{n}\left(\frac{t_i}{(1+q)^{i-1}}\right)^{\beta} = 0$$

Thus, the first equation is

$$\alpha = \left( \frac{\sum_{i=1}^{n}\left(\frac{t_i}{(1+q)^{i-1}}\right)^{\beta}}{n} \right)^{\frac{1}{\beta}} \tag{17-1}$$

Taking the derivative with respect to $\beta$ one gets

$$\frac{d \ln(L(\alpha,\beta,q))}{d\beta} = \frac{1}{\beta} + \ln\left(\frac{t_1}{\alpha}\right) - \left(\frac{t_1}{\alpha}\right)^{\beta} \ln\left(\frac{t_1}{\alpha}\right) + \frac{1}{\beta} + \ln\left(\frac{t_2}{\alpha(1+q)}\right) - \left(\frac{t_2}{\alpha(1+q)}\right)^{\beta} \ln\left(\frac{t_2}{\alpha(1+q)}\right) + \frac{1}{\beta} + \ln\left(\frac{t_3}{\alpha(1+q)^2}\right) -$$

$$- \left(\frac{t_3}{\alpha(1+q)^2}\right)^{\beta} \ln\left(\frac{t_3}{\alpha(1+q)^2}\right) + \ldots + \frac{1}{\beta} + \ln\left(\frac{t_n}{\alpha(1+q)^{n-1}}\right) - \left(\frac{t_n}{\alpha(1+q)^{n-1}}\right)^{\beta} \ln\left(\frac{t_n}{\alpha(1+q)^{n-1}}\right)$$

$$= \frac{n}{\beta} + \sum_{i=1}^{n} \ln\left(\frac{t_i}{a(1+q)^{i-1}}\right)\left[1 - \left(\frac{t_i}{a(1+q)^{i-1}}\right)^{\beta}\right] = 0$$

Accordingly, the second equation is

$$\beta = \frac{n}{\sum_{i=1}^{n} \ln\left(\frac{t_i}{a(1+q)^{i-1}}\right)\left[\left(\frac{t_i}{a(1+q)^{i-1}}\right)^{\beta} - 1\right]} \tag{17-2}$$

And taking the derivative with respect to $q$ one gets the third equation

$$\frac{d \ln(L(\alpha,\beta,q))}{dq} = 0,$$

which is

$$\beta \sum_{i=1}^{n} \frac{i-1}{1+q} \left[\left(\frac{t_i}{a}\right)^{\beta}(1+q)^{\beta(1-i)} - 1\right] = 0. \tag{17-3}$$

Again, Equations (17.1–3) can be solved numerically to obtain MLE estimates of the G1R Process with the underlying Weibull distribution.

*3.4. Case Study*

Consider failure times between 12 consecutive failures discussed by Basu & Rigdon (2000): {3, 6, 11, 5, 16, 9, 19, 22, 37, 23, 31, 45}. The data are of the failure-terminated type. The G1-Renewal process with the underlying exponential distribution is assumed as a probabilistic model. Figure 7 shows MLE of the CIF obtained by solving System (13). It is interesting to note that the CIF exhibits pronounced *convexity*, contrary to *linearity*, which might be intuitively expected from a point process with the underlying exponential distribution. The exponential distribution parameter is estimated (using MLE) to be 4.781 and G1-R restoration parameter as 0.232.

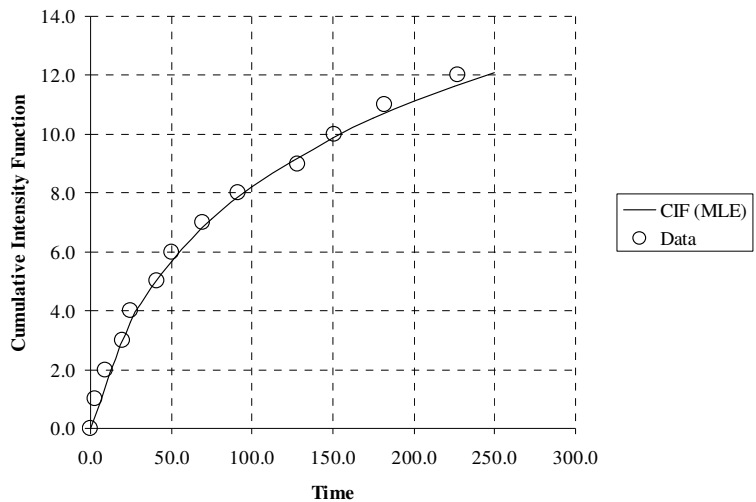

*Fig. 7. G1-Renewal with Exponential Underlying distribution as a Model to Data Set of Basu & Rigdon (2000).*

## About the authors

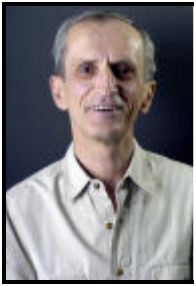

**Mark Kaminskiy** is the Chief Statistician at the Center of Technology and Systems Management of the University of Maryland (College Park), USA. Dr. Kaminskiy is a researcher and consultant in reliability engineering, life data analysis and risk analysis. He has conducted numerous research and consulting projects funded by the government and industrial companies such as Department of Transportation, Coast Guards, Army Corps of Engineers, US Navy, Nuclear Regulatory Commission, American Society of Mechanical Engineers, Ford Motor Company, Qualcomm, General Dynamics, and several other engineering companies. He taught several graduate courses on Reliability Engineering at the University of Maryland. Dr. Kaminskiy is the author and co–author of over 50 publications in journals, conference proceedings, and reports.

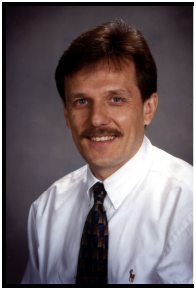

**Vasiliy Krivtsov** is a Senior Staff Technical Specialist in reliability and statistical analysis with Ford Motor Co. He holds M.S. and Ph.D. degrees in EE from Kharkov Polytechnic Institute, Ukraine and a Ph.D. in Reliability Engineering from the University of Maryland, USA. Dr. Krivtsov is the author and co–author of 50+ professional publications, including a book on *Reliability Engineering and Risk Analysis*, 9 patented inventions and 3 Ford corporate secret inventions. He is an editor of the Elsveir's *Reliability Engineering and System Safety* journal and is a member of the IEEE Reliability Society. Prior to Ford, Krivtsov held the position of Associate Professor of Electrical Engineering in Ukraine, and that of a Research Affiliate at the University of Maryland Center for Reliability Engineering. Further information on Dr. Krivtsov's professional activity is available at www.krivtsov.net